\documentstyle[prb,aps,epsf,floats]{revtex}
\begin{document} 

\draft 

\title{Transition Temperature and  Magnetoresistance
in Double-Exchange Compounds with Moderate Disorder} 
 
\author{E.~E.~Narimanov, C.~M.~Varma } 
 
\address{ Bell Laboratories-- Lucent Technologies, 
700 Mountain Ave., Murray Hill NJ 07974}

\date{ \today} 
 
\maketitle 
 
\begin{abstract} 
 We 
develop a variational mean-field theory of the
 ferromagnetic transition in compounds like Lanthanum-Manganite
within the framework of the Double-Exchange Model supplemented by
modest disorder. We obtain analytical expressions for the 
transition temperature, its variation with the 
valence electron-density and its decrease with disorder.  
 We derive 
an expression for the conductivity for both the paramagnetic and the 
ferromangetic metallic phases, and study its
dependence on the temperature and magnetic field. A simple relation between 
the resistivity in the ferromagnetic phase and the spontaneous magnetization is
found. Our
results are in a good agreement with the experimental
data on transition temperatures and resistivity in the manganite compounds 
with relatively small disorder.
We comment on the effects of increased disorder.
\end{abstract} 
\vspace*{-0.05 truein} 
\vspace*{-0.15 truein}

\section{Introduction} 
\label{sec:Intro} 
 
Interest has revived recently in the perovskite manganese 
oxides ${\rm A_{1-x} B_x  Mn O_3}$ (where A is a 
trivalent and B is a trivalent atom), which were first investigated 
in the 1950's.\cite{jonkers} As the doping $x$ and the temperature 
$T$ are varied, these  
manganese oxides show a rich variety of phases\cite{Ramirez}. 
Particularly interesting is the doping region $0.1 \lesssim x  
\lesssim 0.3$, where the compounds 
 undergo a transition from either insulating or  
very high resistance metallic, paramagnetic phase at high  
temperatures  to a ferromagnetic phase at low temperatures. Near the 
transition,  the resistivity of 
the compounds changes by orders of magnitude. The application of a  
strong magnetic field substantially reduces this effect, thus 
giving rise to a very large negative magnetoresistance. The physical 
mechanism, responsible for this behaviour, has been recently the subject 
of much discussion and controversy.  
It was initally suggested\cite{first_de_in_cmr}, that the CMR in manganese 
oxides can be explained within the framework of the Double-Exchange  
Model\cite{Zener} (DE). In this model it is assumed that the on-site 
direct repulsion $U$ is the largest energy, followed in order by 
the Hund's rule  
energy $J$ and the  hybridization energy $t$ between Mn-orbitals 
at neighboring sites. The basic conduction step is then the 
interchange of valence between 
neighboring Mn : $[{\rm Mn^{+3}Mn^{+4} \rightarrow Mn^{+4}Mn^{+3}}]$.  
The basic physical idea of the DE mechanism is  
that this electron 
conduction  
is largest when the initial and the final states are 
degenerate. The latter requirement corresponds to an alignment of the  
spins of the manganese ions. In the opposite case, the conduction rate  
is suppressed 
by a factor of $t/J$.  As a result,  
a transition  
from a paramagnetic to a ferromagnetic 
state leads to a dramatic increase of the conductivity of the compound. 
Using a Dynamical Mean Field calculation (DMFT),\cite{Georges96} the 
double-exchange mechanism was successfully used \cite{Tokura} for 
a quantitative description of the experimental data in ${\rm La Sr Mn O_3}$ 
compounds. A later study \cite{Edwards} claimed that the  
agreement with the experiment found in Ref. \onlinecite{Tokura} was  
caused by  
an unphysical choise of the density of states (DOS), and was accidental. But 
a calculation by Furukawa \cite{Furukawa} with several different  
choises  
of the local DOS confirmed the results of his earlier work\cite{Tokura}. 
 
Subsequently a  calculation carried out by the authors of Ref.  
\onlinecite{Millis1}, concluded that the double exchange model 
alone could not explain the experimental data for the manganese oxides. 
There were two objections: (i) that the double-exchange model gave a 
transition temperature an order of magnitude larger than experiments 
and (ii) that the often observed insulating-like resistivity 
(resistivity increasing with decreasing temperature) could not be 
explained by the double-exchange model. 
It was proposed in Ref. \onlinecite{Millis1}, that for the description 
of mangenese oxides, one should take into account a continuation to 
the metallic state of the  Jahn-Teller distortion found for the 
insulating antiferromagnetic end-member ($x\approx 0$) in these 
compounds into the $x$-range of interest min some kind of dynamic fashion. 
As shown by a simple calculation \cite{Varma96}, objection (i) turns 
out to be due to an inadequate appreciation of the 
energetics of the double-exchange process. The transition temperature
is related to the difference in the electronic cohesive energy of the
ferromagnetic and paramagnetic phases, and is not given by the
transition temperature of a spin model as in Ref. \onlinecite{Millis1}.
As regards the 
proposal of the effects of a possible Jahn-Teller distortion, 
substantial theoretical effort\cite{Millis2,Alexandrov}
has failed to produce any results which can be compared to experiments for the 
resistivity. 
 
Meanwhile, there has been further progress experimentally. It was only 
recently pointed out\cite{Furukawa}, that the manganese 
oxides at similar elecron densities 
show two qualitatively different types of behaviour: 
(i) a metal-insulator transition near $T_c$, which in this case has relatively 
low values $\sim 280$ K and (ii) a metallic behaviour both below  
(a good metal) and above (incoherent metal with the absolute  
value of the resistivity near the Mott's limit) the critical  
temperature, which is comparably high ($\sim 380$ K).\cite{examples}
The difference appears to be the amount of disorder.
This would tend to remove the 
possibility that the Jahn-Teller effects, were they to occur, have much to 
do with the resistivity behavior. Instead the question to ask is why
disorder so dramatically modifies the temperature dependence of
the conductivity in the paramagnetic phase, while simultaneously
reducing the transition temperature. The relation of the resitivity to the 
magnetization $M(T)$ in the
 ferromagnetic phase also
depends on disorder. For small disorder the temperature dependent part is proportional
to the $M(T)^2$, while for large disorder a much stronger dependence is
found.
  
It was suggested\cite{Varma96} that disorder effects-due to spin-disorder, 
lattice polarons  
due to the 30\% difference in the volume of the ${\rm Mn^{+3}}$ and the 
${\rm Mn^{+4}}$ 
ions as well as extrinsic disorder acting in concert might be responsible for 
the resisitivity in the paramangetic phase. The possibility suggested that 
spin-disorder alone my be sufficient turns out, as shown by recent 
numerical studies, not to be correct.\cite{ShengPRB,Li,ShengPRL} Additional 
randomness due to subsitution disorder has been used in calculations to 
explain the experimental data\cite{ShengPRB}. 
Isoelectronic ${\rm La_{0.7-x} R_x Ca_{0.3} Mn O_3}$ shows enormous
descrease of the critical temperature\cite{Hwang95}
when ${\rm R}$ is ${\rm Y}$ compared to
when ${\rm R}$ is ${\rm Pr}$. Note that the ionic radius
of ${\rm La^{3+}}$ is $1.02 \ {\rm A}$, of ${\rm Pr^{3+}}$ is 
$1.01 \ {\rm A}$, and of  ${\rm Y^{3+}}$ is  $0.89 \ {\rm A}$.
Note, that the substituion with ${\rm Y}$ besides changing the average
bond angle introduces disorder.

Also very interesting is the fact that not only does spin-disorder 
disappear for $T \ll T_c$, lattice disorder does as 
well\cite{Shimomura99,Vasiliu99}. This is 
evidenced by the remarkable variation of the Debye-Waller factor 
with temperature below and above $T_c$. It is clear that 
spin and lattice disorder act in concert and quite unusual ways.
Further that quenched lattice disorder generates extra lattice
disorder which is annealed in the ferromagnetic phase.

If indeed the difference in the properties of the CMR materials
is caused by the effect of the substitutional disorder, then it might
be possible to account for the 
main features of the behavior of the ``paramagnetic-metallic'' compounds
using  the ``pure'' double-exchange model.  To address this questions one 
of the
main objectives of the present paper.  We also consider the 
effect of the substitutional disorder, and show that it leads
to a substantial decrease of the critical temperature of the para-
to ferro-magnetic transition, in agreement with the observed difference
in $T_c$ in different CMR materials. In a future paper we hope to address
the more subtle issues connected with cation and other disorder in the
mixed-valent compounds.

The paper is organized as follows. In the next section, we 
develop the variation mean field theory for the double-echange Hamiltonian.
This is a systematization of the ansatz used in ref.\onlinecite{Varma96}. 
We calculate the spin distribution function, and the critical temperature
of the ferromagnetic transtition. In the third section, we study the
effect of the substitution disorder on this phase transition. In 
Section IV we develop a semiclassical transport theory for the
CMR compounds, and calculate the magnetic field- and temperature- 
dependence of the resistivity. We close with a summary and discussion of
future directions.

\section{The Variational Mean Field Theory}
\label{sec:VarMF} 
 
In the semiclassical limit of large spin $S$ of the manganese ions, 
the effective electron Hamiltonian in the double-exchange model   
can be expressed as\cite{Anderson}   
\begin{eqnarray} 
H_{\rm eff} & = & \frac{1}{2} {\sum_{\langle ij \rangle}}' t_0  
\cos(\theta_{ij}/2) c_i^\dagger 
c_j +  {\sum_{ i}} \left[v_i c_i^\dagger c_i 
- \mu_B S B\cos\vartheta_i\right]
\label{eq:H} 
\end{eqnarray} 
where the first sum includes hopping only between the nearest-neighbour 
manganese ions of different valencies, the angle $\theta_{ij}$ is  
defined as the angle between the ion spins ${\bf S}_i$ and ${\bf S}_j$,  
$v_i$ represents the effect of the substitutional disorder, 
$B$ is the magnetic field, and $\mu_B$ is the Bohr magneton. 
The angle $\vartheta$ is the angle  
between the spin  ${\bf S}_i$ and the magnetic field. It is important to 
note that the assumption of large $U$ and $J$ compared to $t$ makes the  
charge carriers effectively spin-less. 
 
Neglecting the correlations in the orientations of the neighbour spins, we  
represent the free energy $F$ of the system in terms of the single spin  
orientation  
distribution function $P_\Omega\left({\bf \Omega}\right)$.  In the 
mean-field approximation, the distribution 
function depends only on the angle $\vartheta$ between the local spin and  
the external magnetic field $B$: 
\begin{eqnarray} 
P_\Omega & = & \frac{1}{2 \pi} \ P_\vartheta\left(\vartheta\right) 
\end{eqnarray} 
In the semiclassical limit, the corresponding spin entropy is then 
\begin{eqnarray} 
S_{\rm spins} & = &   - \int d\vartheta \ \sin\vartheta \  P_\vartheta\left(\vartheta\right) 
\log\left[P_\vartheta\left(\vartheta\right)\right] +  
S^0_{\rm spins}\left(S\right) 
\label{eq:s} 
\end{eqnarray} 
where the function $S^0_{\rm spins}\left(S\right)$ does not depend 
on $P_\vartheta$, and is related to our choise of the  normalization 
of the spin distribution function $\int d(\cos\vartheta) \ P_\vartheta = 1$. 
This semiclassical approximation is discussed in detail in  
Appendix \ref{sec:AppA}. 
 
The  calculation of the energy for a given spin distribution is more  
complicated. When the transfer integral between the 
neighbouring sites $i$ and $j$ is equal to a constant value  
 $\tilde{t}$, and the effects of the substitution disorder can be  
ignored, the electron energy is given by 
\begin{eqnarray} 
E_t\left[\tilde{t}\right] & = &  
\int_{-\infty}^\mu d \varepsilon \ \rho_0\left(\tilde{t};  
\varepsilon\right) \ \varepsilon 
\label{eq:e_t} 
\end{eqnarray} 
where $\rho$ is the electron density of states (DOS) corresponding to the 
Hamiltonian (\ref{eq:H}) with no diagonal disorder ($v_i = 0$) and
constant transfer intergral $t_{ij} = t$. 
To account for the effects of the substitution disorder $v_i$, we introduce 
an effective averaged DOS defined as  
\begin{eqnarray} 
\rho\left(t, \varepsilon\right) = \langle 
\rho_0\left(t, \varepsilon - v\right) \rangle_v 
\end{eqnarray} 
where the average is performed over the distribution of the  
$v_i$'s. 
 
To obtain the total energy, in  the mean field approximation we  
average $E_t[t]$ over the distribution of the transfer integrals 
\begin{eqnarray} 
E & = & \int dt \ P_t\left(t\right) E_t\left[t\right] 
\label{eq:e} 
\end{eqnarray} 
 
\begin{figure}[t] 
\begin{center} 
\leavevmode 
\epsfysize = 6. true cm 
\epsfbox{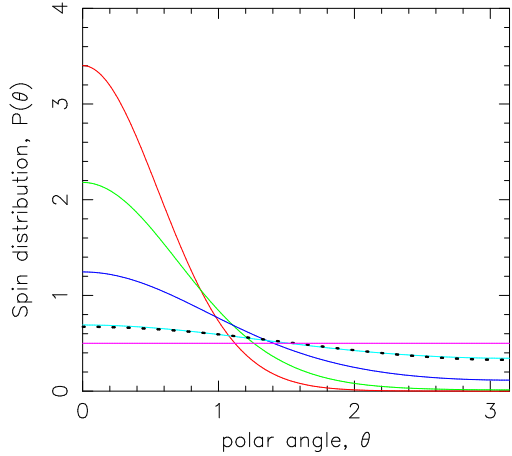}
\end{center} 
\caption{ 
The distribution $P_\vartheta$, for $T/T_c = 0.5$ (red),
$0.7$ (green), $0.9$ (blue), $0.99$ (cyan), $1.01$
(magenta). The dotted line represent the ``linear''
approximation of Eq. (\protect\ref{eq:delta_p}),
appropriate for a small magnetization.
\protect\label{fig:distribution} 
} 
\end{figure}

The transfer integral $t_{ij}$ can then be expressed in terms of the  
polar angles  
$\left(\phi_i, \vartheta_i\right)$ and  
$\left(\phi_j, \vartheta_j\right)$, which define the orientations 
of the corresponding spins, since they uniquely define the relative angle 
$\theta$. Therefore the integration over $t$ in Eq. (\ref{eq:e}) can be  
converted  to the integration over the polar angles. Using the 
 procedure discussed in detail in Appendix
\ref{sec:AppB}, we derive the effective free energy functional, and
by a direct minimization obtain the following integral equation
for the spin distribution function:
\begin{eqnarray} 
P_\vartheta\left(\vartheta\right) & = &  
\exp\left[  
- 2  
\int_0^{2 \pi} \frac{d \phi_1}{2 \pi} 
\int_0^{2 \pi} \frac{d \phi_2}{2 \pi} \ 
\int_0^{\pi} d\vartheta_1 \sin\vartheta_1  
P_\vartheta\left(\vartheta_1\right) 
\int_{-\infty}^\mu d\varepsilon  
\frac{\varepsilon - \mu}{T}  
\rho\left(t_0  
\cos\left(\frac{\theta}{2} 
\right); 
\varepsilon\right) 
- \frac{\zeta}{T} + \frac{\mu_B S B}{T} \cos\vartheta \right] 
\label{eq:p} 
\end{eqnarray} 
Here the parameter $\zeta = \zeta\left(T,B\right)$ accounts for the proper
normalization of the distribution function $P_\vartheta$. The last term
accounts for the energy $\mu_B S B \cos\vartheta$ of the spin,
tilted at the angle $\vartheta$ with respect to the direction of the 
external magnetic field. Finally, the first term in the exponential
of Eq. (\ref{eq:p}) represent the energy of the electron gas, which depends
on the spin distribution via the effective ``local'' bandwidth 
$W \sim \cos\theta/2$, determined by the relative orientation of the
near spins. Note, that this term depends nontrivially
on $\vartheta$ via the relative angle  
$\theta = \theta\left(\phi_1, \vartheta_1; \phi, \vartheta\right)$. 
 This nonlinear integral equation allows a straightforward numerical solution 
by iterations. In Fig. \ref{fig:distribution} we plot the distribution 
$P_\vartheta$ for different values of the scaled temperature  
$\tau \equiv T/t_0$ and dimensionless magnetic field $b \equiv B/t_0$. 
 
When the magnetization of the system is small, and 
the spin distribution is  
close to uniform (e.g. when the system is 
in the paramagnetic phase in  
a small external field), then the distribution function 
\begin{eqnarray} 
P_\vartheta\left(\vartheta\right)  
& = & \frac{1}{2} + \delta p_\vartheta\left(\vartheta\right), 
\ \ \  
\delta p_\vartheta \ll 1 
\end{eqnarray} 
Expanding  the exponential in the right hand side of Eq.  
(\ref{eq:p}) in $\delta p_\vartheta$, and keeping the 
terms  up to the first order in $\delta p_\vartheta$, yields 
\begin{eqnarray} 
\delta p_\vartheta & = & \frac{3}{2} 
{M\left(T,B\right)} 
\cos\vartheta
+ {\cal O}\left(M^2\right),
\label{eq:delta_p} 
\end{eqnarray}
where $M\left(T,B\right) = \chi\left(T\right) B + {\cal O}\left(B^2\right)$ 
is the magnetization of the system. The susceptibility 
$\chi$ is then given by
\begin{eqnarray}
\chi\left(T\right) = \frac{1}{3} \frac{\mu_B^2 S^2}{T-T_c} 
\label{eq:chi} 
\end{eqnarray} 
where the critical temperature $T_c$ is given by 
\begin{eqnarray} 
T_c & = & \int_0^\pi d\vartheta \sin\vartheta \cos\vartheta 
 \int_{-\infty}^\mu  
d\varepsilon \ 
\left(\mu - \varepsilon\right) \rho\left(t_0 \cos\frac{\vartheta}{2} ;  
\varepsilon \right)
\label{eq:t_c} 
\end{eqnarray}
 
It is worthwhile noting the relation of this theory to an earlier 
mean-field variational description of the manganese oxides of 
Ref. \onlinecite{Varma96}. There, in addition to the mean-field 
approximation, a specific functional form of 
the probability distribution of the angle between different spins 
was assumed, with the system magnetization being the variational 
parameter. This should be contrasted to the method of the present  
paper, when the functional form of the single-spin 
distribution function is derived variationally. The general dependence  
of the distribution derived here turns out to be quite similar to 
the one assumed earlier. But the results of the variational procedure developed 
in this section should be more accurate besides being on firmer ground.  
Another advantage of the 
present method is that it can be used for the description of the 
effects of the substitution disorder - something, which is hard 
to characterize within the framework of Ref. \onlinecite{Varma96}. 
 
\begin{figure}[t] 
\begin{center} 
\leavevmode 
\epsfysize = 6 truecm 
\epsfbox{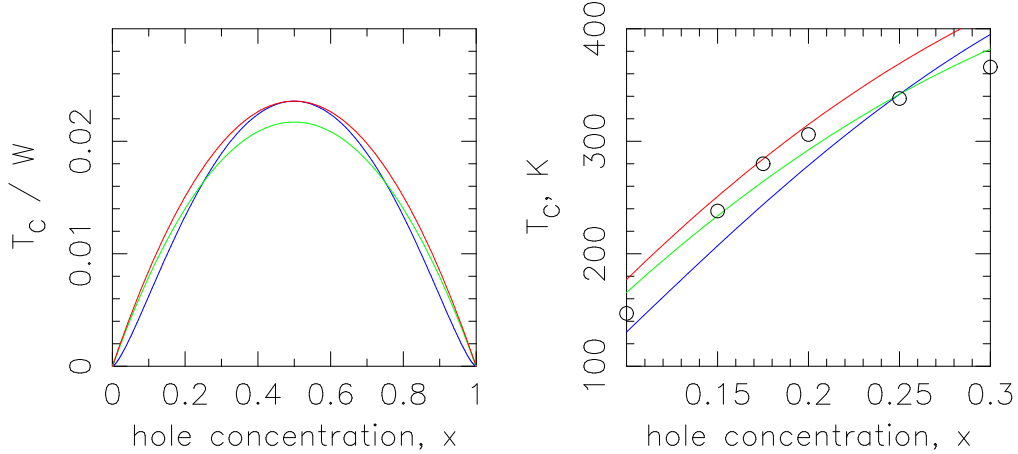}
\end{center} 
\caption{ 
The dependence of the critical temperature on the electron  
concentration. The blue and green lines correspond to respectively rectangular 
and Gaussian DOS. The red line represents the $x(1-x)$ dependence 
obtained in Ref. \protect\onlinecite{Varma96}. 
The circles are the experimantal 
data of Ref. \protect\onlinecite{Tokura}. The left panel shows $T_c$ in 
units of bandwidth $W$, while the absolute units for the right panel were 
calculated, assuming $W = 1.8{\rm eV}$.  
\label{fig:t_c} 
} 
\end{figure}

\begin{figure}[t] 
\begin{center} 
\leavevmode 
\epsfysize = 6. truecm 
\epsfbox{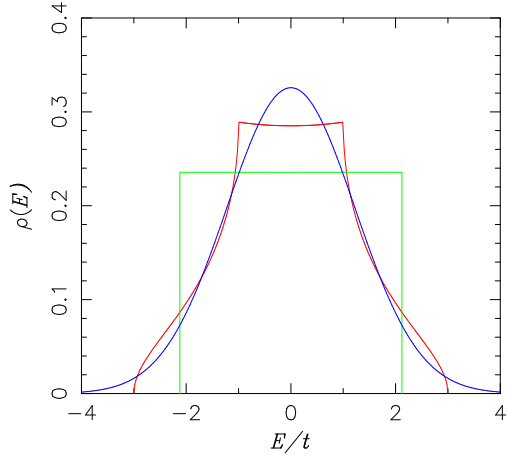} 
\end{center} 
\caption{ 
The model rectangular (green) and Gaussian (blue) densities of states, and
the actual DOS 
$\protect\rho_t\protect\left(\protect\varepsilon\protect\right)$.
For each model density of states, the bandwidth is chosen such that
the second moment $\protect\langle \protect\varepsilon^2 \protect\rangle$
is exact. Note, how accurately the Gaussian DOS fits the 
profile of $\protect\rho_t$.
\label{fig:dos} 
} 
\end{figure}

As follows from Eq. (\ref{eq:t_c}), the critical temperature explicitly  
depends on the  
density of states, and the resulting value is in fact sensitive to  
the actual shape of DOS.
 However, this has only a marginal effect on the 
dependence of $T_c$ on the concentration $x$. To illustrate this 
behaviour, in Fig.  \ref{fig:t_c} we plot the critical 
temperature as a function of the charge carrier concentration $x$ for 
a rectangular (blue curve) and Gassian (red curve) densities of states.   
For comparison, we also plot the $x (1 - x)$ dependence (black line), 
obtained in an earlier work\cite{Varma96}, and the experimental data 
of Ref. \onlinecite{Tokura}. The model densities 
of states are plotted in Fig. \ref{fig:dos} and compared to the
DOS $\rho_t$, corresonding to the hamiltonian (\ref{eq:H}) with constant 
transfer integral, and no diagonal disorder. The effective bandwidth of the 
model desities of states is chosen to accurately reproduce the second
moment $\langle \varepsilon^2 \rangle$. Note, how accurately the 
Gaussian density of states fits the profile of $\rho_t$.
 
As  follows from Fig. \ref{fig:t_c}, a 
reasonable choise of the bandwidth $W = 1.8{\rm eV}$ 
consistent with the calculations in the local density 
approximation\cite{Mattheiss}, leads to a good 
agreement with the experimental data. 
As explained earlier\cite{Varma96}, the transition temperature is determined  
essentially by the difference in the cohesive energy of the ferromagnet  
and the paramagnet by the entropy of the paramagnet. The larger bandwidth of 
the ferromagnet by about 20\% is the essential aspect of the energetics in the 
double-exchange problem. 
 
Consider now the effect of substitutional disorder. Substitutional disorder  
increases the electron-bandwidth for the paramagnet. The removal of 
spin-disorder is then expected to decrease the change in the bandwidth  
on becoming a ferromagnet. This is explicitly borne out by the theory here. 
 
\begin{figure}[t]
\begin{center}
\leavevmode
\epsfysize = 6 truecm 
\epsfbox{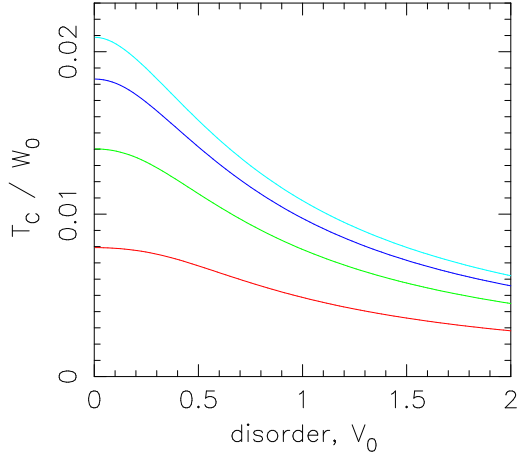} 
\end{center}\caption{
The dependence of the critical temperature on the average ``disorder''
$v_0$ (in units of the bandwidth $W_0$ of the ``clean'' system $v \equiv 0$). 
Curves of different colour represent
different electron concentrations (from top to bottom: $x = 0.4$,
$0.3$, $0.2$, $0.1$). The distribution of the disorder
energies $v_i$ is Gaussian.
\label{fig:tc_disord}
 }
\end{figure}

Since the critical temperature is directly related to the effective DOS, it 
is sensitive to the substitution disorder in the system. 
Assuming the Gaussian distribution of the disorder strength $v_i$ with the 
standard deviation ${V_0}$ and Gaussian ``bare'' DOS\cite{explain_3} 
$\rho_g \propto \exp\left[ - \varepsilon^2/\left(3 t^2\right)\right]$, 
we obtain: 
\begin{eqnarray} 
T_c & = & \int_0^\pi d\vartheta \int_{-\infty}^\mu
d\varepsilon \   
\left(\mu - \varepsilon\right) 
\rho\left[t_{\rm eff}\left(\vartheta\right), \varepsilon \right] \
\sin\vartheta \cos\vartheta  
\label{eq:tc_disord} 
\end{eqnarray} 
where the effective transfer integral $t_{\rm eff}$ is defined by the
equation
\begin{eqnarray}
\frac{1}{t_{\rm eff}^2} & = & \frac{1}{t_0^2 \cos^2\frac{\vartheta}{2}}
+ \frac{3}{V_0^2}
\label{eq:t_eff}
\end{eqnarray}
The dependence (\ref{eq:tc_disord}) is shown in Fig. \ref{fig:tc_disord} 
for different electron concentrations. 
As extra disorder makes the ferromagnetic phase less favourable, the  
critical temperature goes down with an increase of $V_0$.

It might be tempting to attribute the difference in critical temperature 
between  
the ``type-I'' and ``type-II'' compounds to the effect of the substitutional 
disorder.
In such model, an effective disorder strength of $V_0 \sim 
0.7 W$, 
would fully account for only $\sim 30 \%$ difference in the critical 
temperatures 
of ``type-I'' and ``type-II'' compounds. We suspect that large enough 
lattice disorder,
in concert with lattice disorder localises electronic states in the 
paramagnetic phase.
New considerations then eneter in to the determination of the transition 
temperature.
These will be discussed separately. Also missing from the discussion above 
is the
effect of the formation of spin-polarons which must occur in the 
paramagnetic phase
\cite{Varma96,Pokrovsky98}. 
They would tend to decrease $T_c$ but the number of spins 
in the polarons is rather small and only a modest numerical effect on the 
transition
temperature is expected. They are however quite important for the dynamics 
near the 
transition.

\section{Resistivity without lattice disorder: Semiclassical Treatment} 
\label{sec:ResSC} 
 
In the mean-field approximation developed in the previous section,  
each spin independently fluctuates around the averaged value defined by the 
magnetization of the system. 
From the point of view of the semiclassical transport theory, that would 
correspond to effective independent ``scatterers'' located at each 
point of the lattice. However, in the ferromagnetic phase, when the spin 
fluctuations are small compared to the averaged value, the corresponding 
electron mean free path may be substantially larger that the (Mn) lattice  
spacing. In this limit, in order to estimate the resistivity of the system, 
we can use the standard semiclassical transport theory. 
 
We introduce the average transfer integral $\bar{t} \equiv  t_0   
\langle \cos\left(\theta_{\alpha \beta}/2\right) \rangle$, so that the
the corresponding unperturbed Hamiltonian is defined as 
\begin{eqnarray} 
H_0 & = & \frac{1}{2}\sum_{<\alpha \beta>} \bar{t} \ c_\alpha^+ c_\beta 
\label{eq:H0} 
\end{eqnarray} 
and rest of $H$ is treated as the ``perturbation''  
\begin{eqnarray} 
V & = & \frac{1}{2} \sum_{<ij>} \delta t_{ij} \ c_i^+ c_j 
\label{eq:V} 
\end{eqnarray} 
The standard plane-wave diagonalization of $H_0$ yields the dispersion 
law 
\begin{eqnarray} 
\epsilon_{\bf k} = \bar{t} \left[ \cos\left(k_x a\right) + 
\cos\left(k_y a\right) + \cos\left(k_z a\right) \right] 
\end{eqnarray} 
which describes ``holes'' near ${\bf k} = 0$ and ``electrons'' near e.g. 
${\bf k} = \pi/a \left(1,1,1\right)$. When the Fermi energy is located 
near  the bottom or near the top of the band, one can define the effective 
mass for the electrons and the holes respectively,  
$m_* = \frac{2 \bar{t} a^2}{\hbar^2}$. 
 
  The kinetic equation for the electron 
distribution function $f_{\bf k}$ is \onlinecite{K&L},  : 
\begin{eqnarray} 
- e {\bf E} \ \frac{\partial f^0}{\partial {\bf k}} = \frac{2 \pi}{\hbar} 
\sum_{{\bf k}} \left| \langle {\bf k} \left| V \right| {\bf k}' 
\rangle \right|^2 \delta\left( \epsilon_{\bf k} -  \epsilon_{{\bf k}'}\right) 
\left( f_{\bf k} - f_{{\bf k}'}\right) 
\label{eq:kinetic_eqn} 
\end{eqnarray} 
where $f^0(\epsilon_{\bf k})$ is the equilibrium distribution  
function, and $\langle {\bf k} \left| V \right| {\bf k}' 
\rangle$ is the matrix element of the ``perturbation'' (\ref{eq:V}). 
 
The kinetic equation (\ref{eq:kinetic_eqn}) has the solution: 
\begin{eqnarray} 
f_{\bf k} = f^0\left(\epsilon_{\bf k}\right) - e  E  
\tau\left(\bf k \right) 
\frac{ \partial \epsilon_{\bf k}}{\partial  k_z} 
\frac{ \partial f^0\left(\epsilon_{\bf k}\right)}{\partial \epsilon_{\bf k}} 
\label{eq:f} 
\end{eqnarray} 
where the relaxation time is defined by the following equation: 
\begin{eqnarray} 
\frac{1}{\tau\left(\bf k \right)} & = & 
\frac{3 a^3}{2 \pi^2 \hbar} \overline{\delta t^2}  
\int d{\bf k}'  
\left(  
1 + \frac{\epsilon_{{\bf k} + {\bf k}'} }{6 \bar{t}}  
\right) 
\delta\left( \epsilon_{\bf k} -  \epsilon_{{\bf k}'}\right) 
\nonumber \\ 
& - &  
\frac{a^3}{2 \pi \hbar} \overline{\delta t^2}  
\int d{\bf k}' 
\sin^2\left(k'_z a\right) 
\delta\left( \epsilon_{\bf k} -  \epsilon_{{\bf k}'}\right) 
\label{eq:tau} 
\end{eqnarray} 
Assuming a {\it uniform} dispersion $\epsilon_{\bf k} = 
\epsilon\left(\left|{\bf k}\right|\right)$, this expression 
reduces to the standard result for the transport relaxation time
\begin{eqnarray} 
\frac{1}{\tau\left({ k} \right)} & = & 
\int d{\bf k}' \  W_{\bf k k'}
\left(1 - \cos\theta_{\bf k k'}\right)
\delta\left( \epsilon_{\bf k} -  \epsilon_{{\bf k}'}\right) 
\end{eqnarray}
where the scattering rate 
$W_{\bf k k'} = 3 a^3 \left(2 \pi^2 \hbar\right)^{-1}
 \overline{\delta t^2} \left(  
1 + \epsilon_{{\bf k} + {\bf k}'} /\left({6 \bar{t}}\right)
\right)$, and 
$\theta_{\bf k k'}$ is the angle between the
vectors ${\bf k}$ and ${\bf k'}$.

Near the top and the bottom of the band, the integrals in  
(\ref{eq:tau})  allow a straightforward analytical  
evaluation. For example, for the holes we obtain: 
\begin{eqnarray} 
\tau^{-1}\left({\bf k}\right) & = &  
\frac{6}{\pi \hbar}  ka \left(1 - \frac{2}{9} 
\left(ka\right)^2\right) 
\frac{\overline{\delta t^2}}{\bar{t}} 
\label{eq:tau_h} 
\end{eqnarray} 
As we pointed out before, the semiclassical approach developed in the 
present section is appropriate only when the charge carrier mean free 
path $\ell \gg a$. Using (\ref{eq:tau_h}), for the ratio of the mean 
free path to the lattice spacing near the top of the band 
we obtain: 
\begin{eqnarray} 
\frac{\ell}{a} & = & 
\frac{\bar{t}^2}{\overline{\delta t^2}} 
\frac{3}{\pi \left( 1 - \frac{2}{9} \left(ka\right)^2\right) } 
\label{eq:ell} 
\end{eqnarray} 
As follows from Eq. (\ref{eq:tau}),(\ref{eq:ell}), 
in the absence of substitution disorder,  
${\ell/a}$ is {\it always} greater than  
$(3/\pi)*(\bar{t}^2/\overline{\delta t^2})$. The ratio  
$\bar{t}^2/\overline{\delta t^2}$ is a monotocially decreasing 
function of temperature in the ferromagnetic phase, and constant above 
the $T_c$, where $\bar{t}^2/\overline{\delta t^2} = 8$. Therefore,  
since the mean free path due to the spin disorder  
is substantially larger than the effective lattice spacing, 
we expect that in the relevant concentration range 
$x \simeq 0.1 - 0.3$ such a ``pure DE'' system would generally  
show the metallic behavior . Indeed, in a typical ``type-II'' compound 
$\rm La_{0.7}Sr_{0.3}MnO_3$,  the resistivity 
does show the metallic behaviour $d\rho/dT > 0$ both above and below the 
transition\cite{Tokura}. 
 
\begin{figure}[t] 
\begin{center} 
\leavevmode 
\epsfxsize = 16truecm 
\epsfbox{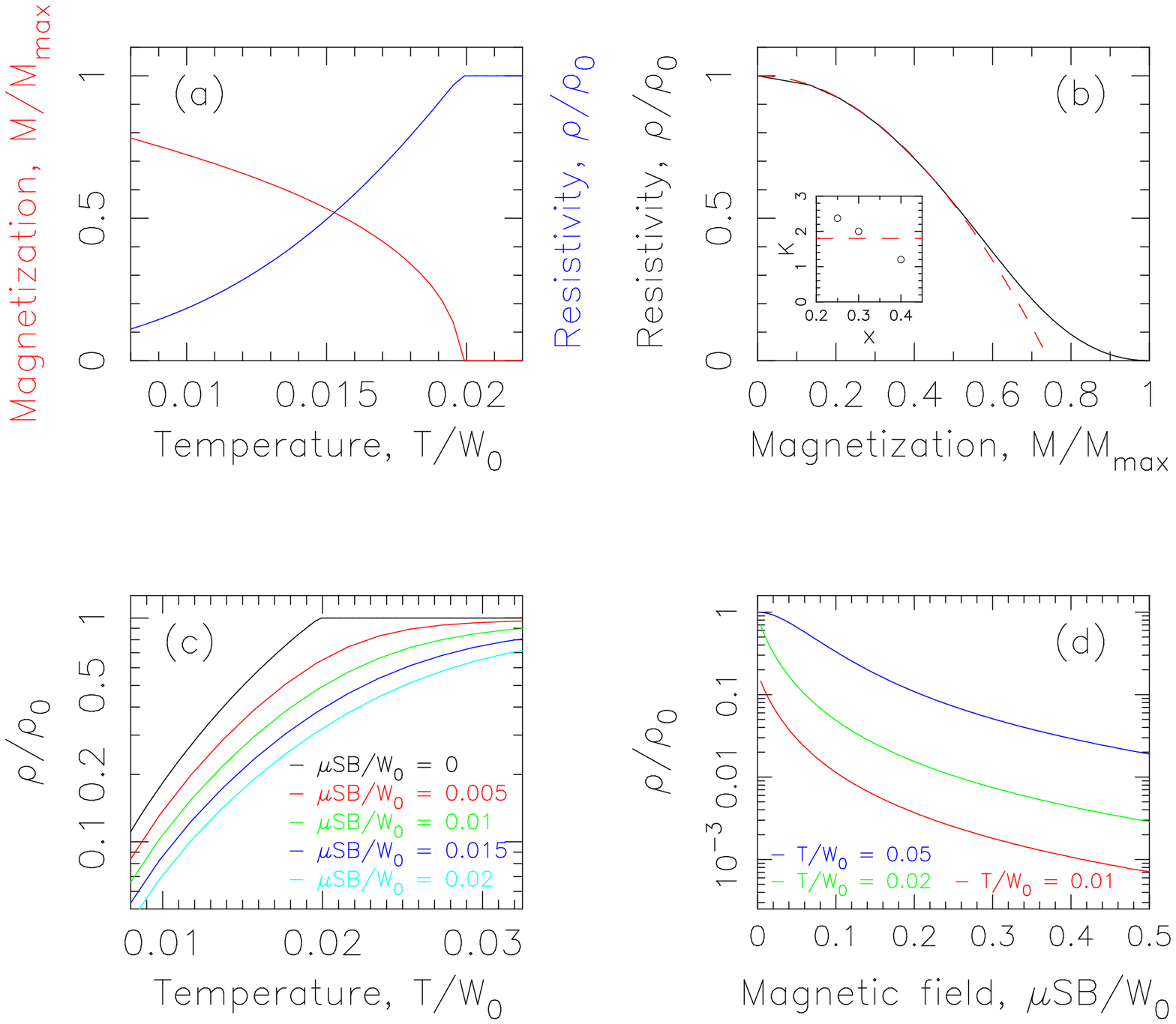} 
\end{center} 
\caption{ The resistivity and magnetization of the sample, calculated
using in the variational mean-field approximation for the DE model.
(a) the vatiation of the resistivity (red line) and 
magnetization (blue line) with temperature, in the absence of 
external magnetic field; (b)   the resistivity as a 
function of magnetization, the inset shows the variation of the 
coefficient $\kappa$ with concentration $x$, obtained from the 
experimental data\protect\cite{Urishibara95}, and the theoretical
prediction in the effective mass approximation (dashed line); 
(c) the resistivity as a function of 
temperature, for different values of the magnetic field;
(d) the resistivity as a function of magnetic field, for different
temperatures.
\label{fig:rho} 
} 
\end{figure}

The effect of nonzero magnetization (caused either by the transtion to the  
ferromagnetic phase, or by external magnetic field) on the conductivity  
is twofold: first, it suppresses the fluctuations in transfer  
integrals thus decreasing the corresponding scattering rate; second, 
the increase of the {\it average} transfer integral caused by the  
magnetization leads to  
a decrease of the effective mass $m_* \sim 1/\bar{t}$. Both these factors 
lead to a decrease of the resistivity $\rho$. For a small magnetization, 
\begin{eqnarray} 
\rho(M) = \rho_0 (1 - \kappa \left(M/M_{\rm max}\right)^2) 
\end{eqnarray} 
where in the effective mass approximation
the coefficient $\kappa$ is equal to $9/5$. For weakly disordered manganites, 
the resitivity indeed follows Eq. (23). As seen in the inset
 in Fig.5(b) the experimental value is
 about 2 and slowly varies with the electron density. Taking into
account the band nonparabolicity leads to a weak dependence of $\kappa$ on
the concentration $x$, but does not fully account for an increase
of $\kappa$.
 The variation of the resistivity in the whole range of the 
sample magnetization $0 < M < M_{\rm max}$ is shown in Fig. \ref{fig:rho}.
 
\section{Substitution disorder: the effect on resistivity}
\label{sec:ResDis} 

As we pointed out in the section \ref{sec:VarMF} of the present paper, the 
$30\%$ difference in the critical temperatures of the ``type-I'' and
``type-II'' compounds implies, that in the ``disordered'' compounds 
the effective scattering potential is of the order of the 
electron bandwidth. In such conditions, the localization effects can
become important, and the semiclassical treatment of the previous 
section is no longer appropriate.
 
It has been proposed, that in the  ``strongly disordered'' 
(``type-I'') compounds, the  ferro- to paramagnetic 
phase transition drives the metal-insulator transition.
In the paramagnetic
phase,
the ``combined effort'' of the substitution and spin disorder
is sufficient to localize the charge carriers, 
while in the ferromagnetic phase, due to larger electron
bandwidth and weaker spin disorder, the mobility edge is below the
Fermi energy\cite{Varma96,ShengPRL}. 

One might be tempted to think that 
this mechanism of the colossal magnetoresistance
of the ``disordered'' manganites reduces the problem to an Anderson-type
transition as a function of disorder alone, where the spin disorder
is a function of the magnetization. This is not correct since the
magnetic  entropy is essential to the transition which 
occurs at a finite temperature unlike the Anderson transition which
occurs at $T=0$.

An important question however is wheather the resistivity near the
transition can be expressed uniquely as a function of magnetization.
If the phase transition (with or without ``diagonal'' disorder)
is characterized by a divergent magnetic correlation length scale,
this may be possible. It should be remembered however that resistivity
depends on fluctuations at large momentum transfer. In a solid with 
lattice disorder, the ferromagnetic correlation length does not
uniquely characterize the important disorder at short length scales
even though it may be coupled to the magnetization as appears to be the
case in the manganites. It is also possible that for sufficiently
strong disorder, the ferromagnetic transition is replaced by a 
cross-over and there is no divergent correlation length. These are
probably the reasons why no clear indication of scaling 
behavior expected in a continuous quantum phase transition
were found in the recent esistivity measurements\cite{Smolyaninova99}.
These considerations, however, go beyond the mean-field type theory,
developed in the present paper.

In any case, we believe that in order to fully understand the physics 
underlying the
colossal magnetoresistance in doped mangnites, one also has to take into
account the lattice disorder, which is coupled to the spin disorder, via
their influence on the charge carriers. Indeed, the strong coupling
between the spin and lattice disorder was recently demonstrated in
two independent experiments.\cite{Shimomura99,Vasiliu99} 
Clearly, the effective lattice disorder has it's own
nontrivial temperature dependence, and, being coupled to the charge 
carriers, therefore obviously leads to 
substantial deviations from the standard picture of the
``static'' Anderson metal-insulator transition. However, at this point
we defer the further description of this effect.

\section{Conclusions} 
 
As shown in the first part of the present paper, the phase  
transition from paramagnetic to ferromagnetic phase in relatively
pure manganese oxides 
can be successfully described by a variational 
approximation on the double-exchange Hamiltonian. The  results
 obtained for the critical  
temperature and its evolution with doping and the chemical composition  
of the compound are consistent with the experimental data. The decrease of
$T_c$ with modest disorder is also understood.
 
The resistivity of the ``type-II'' manganese compounds can also be  
successfully described using the DE model. We showed, that e.g. the 
theoretical dependence of the resistivity on the sample magnetization 
is in a quantitative agreement with the expetimental data. Our calculations  
show the robustness of the results to the particular choice of the 
electron density of states, as should be obvious since the transition
 temperature depends on the difference of the cohesive energy of 
the paramagnetic and the ferromagnetic phases.

The principal problems of the manganites left unanswered in this paper
concern the properties of the "type-I" compounds and the remarkable
effects of disorder in them in both the dynamic and static properties.
These are also the more subtle problems.
Especially interesting is the fact that extrinsic disorder appears to promote
some additional disorder in the paramagnetic phase which is swept away
togather with the spin-disorder in the ferromagnetic phase.
We hope to provide an answer to these questions separately.
 
\appendix 
 
\section{The Spin Entropy in the Semiclassical Mean-Field Approximation} 
\label{sec:AppA} 
 
In the mean field approximation, when the spin density matrix 
of the whole system $\rho^{\Sigma}\left(S_z^{(i)}\right)$ 
 is represented as a product of diagonal density matrices 
$\rho_{i} = \rho^{(1)}\left(S_z^{(i)}\right)$ of  
the individual spins  
\begin{eqnarray} 
\rho^{\Sigma} & = & \Pi_{i=1}^N \rho_i 
\end{eqnarray} 
Then the total spin entropy  
\begin{eqnarray} 
S_{\rm spins}^\Sigma & = & {\rm Tr}\left\{\rho^{\Sigma}  
\log\left[\rho^{\Sigma}\right] \right\}
\end{eqnarray} 
is represented by 
\begin{eqnarray} 
S_{\rm spins} = N \sum_{S_z = -S}^{S} \rho^{(1)}\left(S_z\right) 
 \log\left[ \rho^{(1)}\left(S_z \right)   \right] 
\label{eq:S_Sz} 
\end{eqnarray} 
In the semiclassical approximation $S \gg 1$, the summation over 
$S_z$ can be replaced by integration. Introducing new variable 
$\vartheta \equiv \arccos(S_z/S)$, we obtain: 
\begin{eqnarray} 
S_{\rm spins} = N S \int_{-1}^1 d\cos\vartheta \  
\rho^{(1)}\left(S \cos\vartheta\right) 
 \log\left[ \rho^{(1)}\left(S \cos\vartheta \right)   \right] 
\end{eqnarray} 
where $\rho^{(1)}\left(S \cos\vartheta\right)$ is normalized as follows: 
\begin{eqnarray} 
1 & = & \sum_{{S_z} = -S}^S \rho^{(1)}\left(S_z\right) =  
S \int_{-1}^1 d\cos\vartheta \  
\rho^{(1)}\left(S \cos\vartheta\right) 
\label{eq:sz_norm} 
\end{eqnarray} 
We now define the spin orientation distribution function $P_\vartheta 
\sim \rho^{(1)}\left(S \cos\vartheta\right) $, 
normalized as  
\begin{eqnarray} 
 \int_{-1}^1 d\cos\vartheta \  
\rho^{(1)}\left(S \cos\vartheta\right) & = & 1 
\label{eq:theta_norm} 
\end{eqnarray} 
As follows from Eqns. (\ref{eq:sz_norm}), (\ref{eq:theta_norm}), 
the spin orientation distribution function 
\begin{eqnarray} 
P_\vartheta & = & \frac{1}{S} \rho^{(1)}\left(S \cos\vartheta\right) 
\end{eqnarray} 
Therefore, the semiclassical spin entropy 
\begin{eqnarray} 
S_{\rm spins} = - N \int_{-1}^1 d\cos\vartheta \ P_\vartheta 
 \log\left[ P_\vartheta   \right] +  
N \log\left[S\right] 
\label{eq:S_sc} 
\end{eqnarray} 
For example, in the paramgagnetic phase, when there are no external fields, 
and the spind orienataion distribution is uniform, $P_\vartheta = 1/2$,  
the semiclassical spin entropy is equal to $N \log(2S)$, which is 
consistent with the exact result $N \log(2 S + 1)$ for $S\gg1$. Note, that 
the main contribution to the semiclassical spin entropy comes actually 
from the distribution-independent term in Eq. (\ref{eq:S_sc}). 
 
The semiclassical description, however, fails for large magnetization, 
when the spin system is almost completely polarized, and the  
distribution function starts to change substantially on the scale of  
$\delta \vartheta \sim 1/S$. In this case, the original expression, 
Eq. (\ref{eq:S_Sz}), should be used for the calculation of the spin 
entropy.

\section{The Variational Free Energy Functional} 
\label{sec:AppB} 

In the present Apendix, we calculate the variational free energy
functional for the double-exchange model. Using Eqns. 
(\ref{eq:e_t}),(\ref{eq:e}), for the electron energy
we obtain: 
\begin{eqnarray} 
E_e & = & \int_0^{2 \pi} \frac{d \phi_1}{2 \pi}  
 \int_0^{2 \pi} \frac{d \phi_2}{2 \pi} 
\int_0^{\pi} d\vartheta_1 \sin\vartheta_1  
\int_0^{\pi} d\vartheta_2 \sin\vartheta_2 \ 
P_\vartheta\left(\vartheta_1\right) 
P_\vartheta\left(\vartheta_2\right) 
\int_{-\infty}^\mu d\varepsilon \varepsilon  
\rho\left( 
t_0  
\cos\left( 
\frac{\theta\left(\phi_1,\vartheta_1; \phi_2, \vartheta_2\right)}{2} 
\right) ; 
\varepsilon\right)  
\label{eq:e_theta} 
\end{eqnarray} 
while the extra spin energy
\begin{eqnarray}
E_s = - B \int_0^{2 \pi} \frac{d \phi_1}{2 \pi}  
\int_0^{\pi} d\vartheta_1 \sin\vartheta_1 \cos\vartheta_1  
P_\vartheta\left(\vartheta_1\right) 
\label{eq:e_spin}
\end{eqnarray} 
The free energy can be obtained by the  subsituting these 
expressions and the entropy (\ref{eq:s}) 
into the standard definition of the free energy
\begin{eqnarray} 
F & = & E_e + E_s - T S 
\label{eq:fe} 
\end{eqnarray} 
In order to find the single-spin 
distribution $P_\vartheta$, one has to minimize the effective free energy, 
taking into account the constraints of  
normalization. Using the standard Lagrange multiplier  
method, for the  
effective free energy functional we obtain: 
\begin{eqnarray} 
\tilde{F}\left[P_\vartheta; \mu,\lambda, \zeta \right]  
& = & \int_0^{2 \pi} \frac{d \phi_1}{2 \pi}  
\int_0^{2 \pi} \frac{d \phi_2}{2 \pi} 
\int_0^{\pi} d\vartheta_1 \sin\vartheta_1  
\int_0^{\pi} d\vartheta_2 \sin\vartheta_2 
P_\vartheta\left(\vartheta_1\right) 
P_\vartheta\left(\vartheta_2\right) 
\int_{-\infty}^\mu d\varepsilon  
\left(\varepsilon - \lambda\right) 
\rho\left(t_0  
\cos\left(\frac{\theta}{2} 
\right)); 
\varepsilon\right) \nonumber \\  
& + &  
T \int_0^{2 \pi} \frac{d \phi_1}{2 \pi}  
\int_0^{\pi} d\vartheta_1 \sin\vartheta_1  
P_\vartheta\left(\vartheta_1\right) 
\log\left[ P_\vartheta\left(\vartheta_1\right) \right] 
 - B \int_0^{2 \pi} \frac{d \phi_1}{2 \pi}  
\int_0^{\pi} d\vartheta_1 \sin\vartheta_1 \cos\vartheta_1  
P_\vartheta\left(\vartheta_1\right) \nonumber \\ 
& + & \zeta  
\int_0^{\pi} d\vartheta \ \sin\vartheta  \  
p_\vartheta\left(\vartheta\right)  
+ \bar{F}\left[ x,\lambda, \zeta\right] 
\label{eq:f2} 
\end{eqnarray} 
where the ``constant'' $\bar{F}$ represents the spin  
distriibution-independent part of the free energy. Here, the  
Lagrange multiplier $\zeta$ accounts for the normalization 
of the distribution function $P_\vartheta$, while the  
Lagrange multiplier $\lambda$ represents the constraint of having a fixed 
concentration of mobile electrons in the system. 
 
It is straightforward to show by a direct calculation, that at the 
extremum of the functional (\ref{eq:f2}) $\lambda = \mu$. This has 
a clear physical meaning - the Lagrange multiplier $\lambda$ corresponds
to the electron number conservation, and therefore shoul be equal to
the electron electrochemical potential. Replacing $\lambda$ by 
$\mu$ in (\ref{eq:f2}), we finally obtain the 
effective mean field free energy functional: 
\begin{eqnarray} 
\tilde{F}\left[P_\vartheta; \mu,\lambda, \zeta \right]  
& = & \int_0^{2 \pi} \frac{d \phi_1}{2 \pi}  
\int_0^{2 \pi} \frac{d \phi_2}{2 \pi} 
\int_0^{\pi} d\vartheta_1 \sin\vartheta_1  
\int_0^{\pi} d\vartheta_2 \sin\vartheta_2 \ 
P_\vartheta\left(\vartheta_1\right) 
P_\vartheta\left(\vartheta_2\right) 
\int_{-\infty}^\mu d\varepsilon  
\left(\varepsilon - \mu\right) 
\rho\left(t_0  
\cos\left(\frac{\theta}{2} 
\right); 
\varepsilon\right) \nonumber \\ 
& + &  
T \int_0^{2 \pi} \frac{d \phi_1}{2 \pi}  
\int_0^{\pi} d\vartheta_1 \sin\vartheta_1  
P_\vartheta\left(\vartheta_1\right) 
\log\left[ P_\vartheta\left(\vartheta_1\right) \right] 
 - B \int_0^{2 \pi} \frac{d \phi_1}{2 \pi}  
\int_0^{\pi} d\vartheta_1 \sin\vartheta_1 \cos\vartheta_1  
P_\vartheta\left(\vartheta_1\right) \nonumber \\ 
& + & \zeta  
\int_0^{\pi} d\vartheta \ \sin\vartheta  \  
p_\vartheta\left(\vartheta\right)   + \overline{F}
\label{eq:fmf} 
\end{eqnarray} 
Taking the functional derivative of (\ref{eq:fmf}) with  
respect to $P_\vartheta$, for the distribution we obtain: 
\begin{eqnarray} 
P_\vartheta\left(\vartheta\right) & = &  
\exp\left[  
- 2  
\int_0^{2 \pi} \frac{d \phi_1}{2 \pi} 
\int_0^{2 \pi} \frac{d \phi_2}{2 \pi} \ 
\int_0^{\pi} d\vartheta_1 \sin\vartheta_1  
P_\vartheta\left(\vartheta_1\right) 
\int_{-\infty}^\mu d\varepsilon  \ 
\frac{\varepsilon - \mu}{T}  
\rho\left(t_0  
\cos\left(\frac{\theta}{2} 
\right); 
\varepsilon\right) 
- \frac{\zeta}{T} + \frac{B}{T} \cos\vartheta \right] 
\label{eq:p_AppB} 
\end{eqnarray} 
Note, that the exponential on the right hand side nontrivially  
depends on $\vartheta$ via the angle  
$\theta = \theta\left(\phi_1, \vartheta_1; \phi, \vartheta\right)$.


\begin{references}
\bibitem{examples}
A canonical example for the behaviour of the second type is the  
$\rm La_{0.7} Sr_{0.3} Mn O_3$ compound, while the most well-studied system 
of the first group is $\rm La_{0.67} Ca_{0.33} Mn O_3$.
\bibitem{explain_3} The coefficient of $1/3$ is introduced so that 
the Gaussian distribution $\rho_g$ will yield the same second moment
$\langle \varepsilon^2 \rangle$ as the actual density of states - see 
Fig. 4.
\end{references}
\end{document}